\documentclass[aps,floatfix,superscriptaddress,notitlepage,nofootinbib]{revtex4-1}
\usepackage{amsmath,amssymb,amsfonts,graphics,graphicx,dcolumn,bm,enumerate}
\usepackage{comment,natbib,appendix}
\usepackage{multirow,color}
\usepackage{chngpage}
\usepackage{afterpage}
\usepackage{xcolor}
\usepackage{amsthm}
\usepackage{natbib}
\usepackage{hyperref}
\usepackage[margin=1.2in]{geometry}
\usepackage{epstopdf}
\usepackage{float}

\newcommand{\cmp}
{\affiliation{Saha Institute of Nuclear Physics, Kolkata 700064, India.}}
\newcommand{\isi}
{\affiliation{Economic Research Unit, Indian Statistical Institute, Kolkata 700108, India.}}
\newcommand{\raghunathpur}
{\affiliation{Raghunathpur College, Raghunathpur, Purulia 723133, India.}}
\newcommand{\SRM}
{\affiliation{SRM University-AP, Andhra Pradesh - 522502, India.}}

\begin{document}
\title{Success of Social Inequality Measures in
Predicting Critical or Failure Points in
Some Models of Physical Systems}

\author{Asim Ghosh}
\email[Email: ]{asimghosh066@gmail.com}
\raghunathpur
 
\author{Soumyajyoti  Biswas}
\email[Email: ]{soumyajyoti.b@srmap.edu.in}
\SRM 
 
 \author{Bikas K. Chakrabarti }%
 \email[Email: ]{bikask.chakrabarti@saha.ac.in}
 \cmp \isi 

\begin{abstract}
Statistical physicists and social scientists both
study extensively some characteristic  features
of the unequal distributions of energy, cluster
or avalanche sizes and of income, wealth etc
among the particles (or sites) and population respectively. While physicists
concentrate on the self-similar (fractal) structure
(and the characteristic exponents) of the largest
(percolating) cluster  or avalanche, social scientists
study the inequality indices like Gini and Kolkata
etc given by the non-linearity of the Lorenz function
representing the cumulative fraction of the wealth
possessed by different fraction of the population.
We review here, using results from earlier
publications and some new numerical as well as
analytical results, how the above-mentioned social
inequality indices, when extracted from the unequal
distributions of energy (in kinetic exchange models),
cluster sizes (in percolation models) or avalanche
sizes (in self-organized critical or fiber bundle
models) can help in a major way in providing precursor
signals for an approaching critical point or imminent failure
point. Extensive numerical and  some analytical results  have been discussed.
\end{abstract}
 
 \maketitle
 
\section{Introduction}
\label{sec1}
Unequal distributions of resources (for example
income or wealth) among the population are
ubiquitous. Economists, in particular, quantify
such inequalities in distributions using some
inequality indices  (e.g., Gini, Kolkata, etc),
defined through the Lorenz function \cite{inoue1,inoue2}
(see e.g., \cite{ref1}
for recent review). Unequal distributions of energy (per degrees of
freedom), of cluster sizes (sites, bonds, etc), of
avalanches (elements failing in one go), etc in
many-body systems are also ubiquitous and also
extensively studied in various physical systems by
statistical physicists over the ages (see e.g., \cite{ref2,ref3,ref4}).
Physicists usually concentrate on the (fractal)
structure of the biggest (in size) cluster or avalanche, 
which becomes of the order of the system
size, inducing the eventual macroscopic
self-similarity and the consequent critical
behavior characterized by the critical
exponents (see  e.g., \cite{ref3,ref4}).

Economists traditionally quantify the  social
inequalities in distributions using inequality
indices, defined through the Lorenz function
$L(f)$ \cite{ref5}. After ordering the population from
the poorest to the richest, the Lorenz function
$L(f)$ is given by the cumulative wealth fraction
possessed by the fraction $f$ of the population,
staring from the poorest: $L(0) = 0$ and $L(1)
= 1$ (see Fig. 1). If everyone had equal share of
the wealth, $L(f) = f$ would be linear (called the
equality line)  and the old and still most popular
inequality index,  namely the Gini ($g$) index \cite{ref6},
is given by the area between the equality line and
the Lorenz curve, normalized by the entire area
(1/2) below the equality line. Thus, $g = 0$
corresponds to perfect equality  and $g= 1$
corresponds to extreme inequality. Another
recently introduced inequality index, namely the
Kolkata ($k$) index \cite{ref7}, can be defined as the
nontrivial fixed point of the complementary Lorenz
function $ L_c(f) \equiv 1 - L(f)$: $L_c(k) = k$. 
It says that the richest  ($1 - k$) fraction of the population possesses a fraction $k$ of the total wealth
($k = 1/2$ corresponds to perfect
equality and $k$ = 1 corresponds to extreme
inequality). As such, $k$ index quantifies and
generalizes  the century old 80-20 law (corresponding
to $k$ = 0.80) of Pareto \cite{ref8}. Extensive analysis of
social data (see e.g., \cite{ref9,ref10}) indicates that in extremely competitive situations, the indices $k$ and
$g$ become equal in magnitude of about
0.86 (instead of 0.80).

\begin{figure}[H]
\centering
\includegraphics[width=0.7\textwidth]{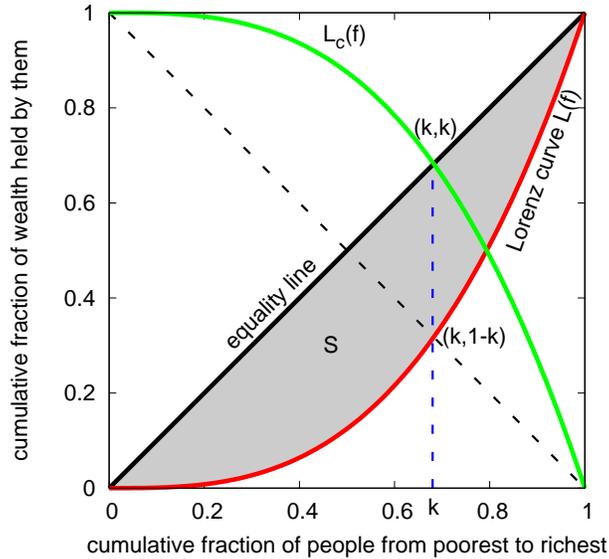}
\caption{The Lorenz curve $L(f)$ (shown in
red) and complementary Lorenz curve $L_c(f)$ (shown
in green) used to calculate the inequality
indices Gini ($g= 2S$, $S$ denoting the area of
the shaded region between the equality line and
the Lorenz curve) and Kolkata (given by the fixed
point $k= L_c(k) \equiv 1 - L(k)$). Here, $L$
represents the cumulative fraction of wealth
possessed by $p$ fraction of the people, when
ordered from poorest to the richest. For the physical systems considered here like
the kinetic exchange model of gas, the wealth
could be replaced by the kinetic energy and the
fraction of people by the fraction of particles.
For  model  systems like percolation, sand piles or
the fiber bundles, the horizontal axis represents
the fraction ($f$) of  clusters or avalanches
when  all avalanches are arranged from lowest to
the highest size. The vertical axis ($L$)
represents the fraction of the cumulative mass of
these  clusters or avalanches.}
\label{fig-1}
\end{figure}

The Gibbs distribution (see e.g., \cite{ref2}) of kinetic
energy among the particles in a classical ideal gas
in equilibrium can also be analyzed in terms of
the corresponding Lorenz function $L((f)$ and then
extracting the Gini ($g$) and Kolkata ($k$) indices
for the kinetic exchange models of market by exploiting the formal similarity between the energy of the gas molecules in the kinetic theory and wealth of an individual and that between temperature and noise in trade (see e.g.,
\cite{ref11,ref12,ref13}). Similarly, the distributions of cluster
sizes (see e.g., \cite{ref4}) in the percolation models
on lattices can be analyzed in terms of the $g$ and
$k$ indices. At occupation concentration $p$
both below and above the percolation threshold
$p_c \simeq 0.5927$ \cite{ref4}  gave $g = k \simeq 0.865$ at a
site occupation probability $p \simeq 0.565$, somewhat below the percolation threshold $p_c$. The statistics of avalanches (successive
failures in one go, without any increase in the
external perturbation), following the  self-organizing
critical dynamics of the sand-pile models (see e.g.,
\cite{ref14}) has also recently been analyzed in terms of the
behavior of social inequality indices $g$ and $k$ \cite{ref15}.
Finally, the avalanche statistics due to collective
dynamics of failure or breaking of individual elements
(in non-brittle materials), studied using the Fiber
Bundle Models (FBMs) (see e.g., \cite{biswas_13,ref16,ref17,ref18}), when analyzed
using the  social inequality indices $g$ and $k$ \cite{ref15,ref19}, gives
intriguing possibility of predicting the imminent
failure of the entire bundle.


Typical FBMs or such failure-prone
(dynamically coupled many-element)
systems, or the percolating systems, are not
self-organized, and are externally driven or
tuned. Particularly for the FBM, a discrete set of elements, each having a failure threshold randomly drawn from a distribution, carries a load. The elements are irreversibly broken when the load on them cross the pre-assigned threshold value. Either the remaining intact elements (fibers) are able to support the applied load, or at a sufficiently high value of the applied load (the critical load for the system), the entire system breaks down. Unlike in the other cases discussed here, there is no dynamics on the other side of the critical point, as the system does not survive at all beyond the critical load.  For Self-Organized-Critical (SOC) systems,
(see e.g., \cite{ref14}), as the average `height' $h$
per site increases, our numerical study (in square
lattice) \cite{ref15} shows that $g$ and $k$ approach
each other and become equal to about 0.863 (BTW),
and 0.856 (Manna) at the respective 
values of average heights $h \simeq  2.087$ (little less than the actual critical value  $h_c=17/8
= 2.125$) for BTW model and $h \simeq 0.6859$
(compared to $h_c\simeq 0.7172$) for Manna model.
For the other SOC models considered here \cite{ref15,ref19}, 
like the
driven-interface Edwards-Wilkinson model and the
centrally pulled FBM show similar growth
(from $g = 0$ and $k = 0.5$) of the inequality
indices to about $g = k \simeq 0.86$ a little below
the respective SOC points.

All these numerical studies indicate that,
except for the irreversible dynamical systems
like FBMs (where the dynamics eventually stops), all  critical systems
(self-organised or otherwise) show a clear
precursor behavior of the inequality indices
like the Gini $g$ and Kolkata $k$. Particularly,
if the inequality in the response of a physical system is measured as it approaches a critical point, such measures show universal trends, irrespective of the universality class of the associated critical point. The particular response to be measured depends on the particular system. For example, in the case of site percolation, it is the inequality between the clusters for a given occupation probability, for kinetic exchange model of wealth, it is the wealth distribution between the individuals, for SOC systems it is the time series of the avalanches. The inequality indices Gini ($g$) and Kolkata ($k$) 
typically start from 0 and 0.5 respectively (for
small and almost equal size clusters) when the systems are away from the critical point. Then they approach
$g = k \simeq 0.86$ slightly before the critical
point is reached. Even for irreversible
systems like the FBMs, the indices $g$
and $k$ assume universal terminal values
about 0.45 and 0.65 respectively, providing
a major statistical precursor signal for
the impending catastrophes.

We will review here some recent numerical studies
on the properties of Gini ($g$) and Kolkata ($k$)
indices for extended kinetic exchange models \cite{ref12},
with some analytical Landau-like formulation of
the Lorenz function $L(f)$ and the analytical
estimates of $g$ and $k$ and the relationships
between them, including an estimate of the
self-organized poverty (energy) level. Next, we
will discuss the numerical observations on
$g$ and $k$ for site percolating system in two
dimensions and discuss, in particular, how their
coincidence in magnitude ($g = k \simeq 0.86$)
occurs preceding the imminent  percolation or 
critical point. Similar results \cite{ref15}
($g = k \simeq 0.86$) as the sand pile systems
approach the  self-organized critical point in
the Bak–Tang–Wiesenfeld sand pile model, Manna
model and a centrally pulled self-organized  fiber bundle model,
will be discussed. Finally, we will discuss the
numerical results  ($g \simeq 0.45$ and $k\simeq 0.65$) as the global breaking point
approach \cite{ref19,ref20} in the equal-load-sharing fiber
bundle models with irreversible local failures
and collective load-share mechanism and their relevance in earthquake statistics \cite{in_prep}.


 \section{ NUMERICAL RESULTS FOR SOCIAL INEQUALITY
INDICES IN KINETIC EXCHANGE, PERCOLATION,
BTW, MANNA \& FIBER BUNDLE MODELS}
\label{sec2}

In this section, we will discuss mostly
numerical results for the Gini \cite{ref1,ref6}
and Kolkata \cite{ref1,ref7} indices for the
kinetic exchange models \cite{ref11,I.1,Ludwig2022},
percolating systems \cite{ref4,Stauffer2017}  and three
Self-Organized-Critical (SOC) models,
namely the Bak–Tang–Wiesenfeld model (BTW) \cite{Bak1987}, Manna Model \cite{Manna1991}  and a self-organizing
centrally-pulled-fiber-bundle model \cite{biswas_13,ref16}.
We also discuss the same for the standard
Fiber Bundle Model (FBM) \cite{Peirce}  (see also
\cite{ref16,ref17,ref18}), where the irreversible breaking
dynamics stops as the whole bundle fails. Except
for the kinetic exchange model considered
here, all the other models exhibit critical,
externally tuned (as in percolation) or
self-organized (as in BTW, Manna, or
centrally pulled fiber bundle) behavior at
(in percolation model) or beyond  (for the
SOC models) the respective critical points
(traditionally identified as the phase
transition point). The bundle failure
points (stress) in such FBMs have already
been identified as the corresponding
critical points \cite{ref16,ref17,ref18,Chakrabarti2021}.

As mentioned earlier, statistical physicists
have studied extensively, over the last five
decades, the building up of self-similarity
in the spatial and temporal structures of the
clusters or avalanches near the critical point,
where the system spanning cluster (corresponding
to the divergent correlation length \cite{ref2,ref3,ref4}) or
the consequent critically slow dynamics
(divergent relaxation time \cite{ref2,ref3,ref4}), characterized
by the (singular and universal) exponents, occur.
This self-similar system-spanning fractal
structure developed at the critical point is
necessarily very much  unequal compared to the
other structures which become quite unimportant
there. The Lorenz function \cite{ref5} of these cluster
or avalanche size distributions near these
critical points 
are found to become such that the Gini ($g$) and
Kolkata ($k$) indices  become equal and nearly
universal ($g = k \simeq 0.86$ or becomes nearly
universal ($g \simeq 0.45, k \simeq 0.65$) at
the breaking or failure points of FBMs. 
This equality ($g=k\simeq0.86$) or otherwise ($g \simeq 0.45, k \simeq 0.65$) will be seen to follow from Monte Carlo data in various model cases discussed in this section and it occurs a little away from the critical point where  the inequalities become much larger.
This universal value of the inequality  indices in various
physical systems, prior to the  arrival of the
respective (widely different) critical
points, can provide excellent precursor
signals.

\subsection{Social indices $g$ \& $k$ in Kinetic Exchange Models}

Let us first recount briefly the derivation of energy
($\epsilon$) distribution $n(\epsilon)$, representing
the number of constituent free (Newtonian) particles
of a classical ideal gas in equilibrium at a
temperature $T$.  If $g(\epsilon)$ denotes the
'density of states’, giving $g(\epsilon)d\epsilon$
equal to the number of dynamical states possible for
any of the free particles of the gas which has a
kinetic energy between $\epsilon$ and $\epsilon +
d\epsilon$ (as counted by the different momentum
vectors $\vec p$ corresponding to the same
kinetic energy $\epsilon \sim |p|^2$ (giving
$g(\epsilon) d\epsilon \sim$
$\sqrt{\epsilon}d\epsilon$ in three dimension),
then one can write $n(\epsilon)d\epsilon
= g(\epsilon) \rho(\epsilon, T)d\epsilon$. For
completely stochastic and ergodic many-body
dynamics or energy conserving exchanges, the
statistical energy distribution function
$\rho (\epsilon, T)$ should  satisfy
$\rho(\epsilon_1 )\rho(\epsilon_2 ) =
\rho(\epsilon_1 + \epsilon_2)$ for any arbitrary
choice of $\epsilon_1$ and $\epsilon_2$. This
suggests $\rho(\epsilon) \sim \exp(-\epsilon/T)$.
These finally give $n(\epsilon) =
\sqrt \epsilon \exp(\epsilon/T)$, where
the factor $T$ can be identified from the
observed knowledge about the equation of state
for the gas.

In a natural extension of this oldest and most
established many-body theory, econophysicists
developed (see e.g., \cite{ref11,I.1}) the Kinetic
Exchange model of trading markets with fixed
total money ($M = N$) exchanged only among
fixed (large) number  ($N$) agents or traders.
Here the money $m_i(t) (M = \sum_i m_i(t)$)
at any time $t$ (measured by the number of
trades or scattering) of the $i$-th  agent (or
`social atom') is identified as the kinetic
energy ($\epsilon$) of the corresponding atom
or particle. In the market, total amount of
money ($M = N)$  remains conserved as no one
can print money or destroy money (will end-up
in jail in both cases). Following the kinetic
theory picture of random kinetic energy
exchanges among the particles in an ideal gas,
the money  exchanges among the agents in the
market here are  considered  to be completely random.
One would therefore again expect, for
any buyer-seller transaction in the market,
$\rho(m_1)\rho(m2_) = \rho(m_1 + m_2)$, where $\rho(m)$
denotes the equilibrium or steady state
distribution of money $m$ among the traders
in the market. This in turn, in a
similar way, suggests $\rho(m) = A\exp (-m/{\Delta})$,
where $A$ and $\Delta$ are constants. Since
$\int \rho(m) dm = N = M = \int m\rho(m) dm$, we get
finally $n(m) = \rho(m) = \exp (-m)$ for the steady
state number of traders with money $m$ in the
market (since there cannot be any equivalent of
the particle momentum vector for the agents, the
equivalent of the density of states $g(m)$ here
is a constant).

\textbf{a)} One can easily calculate \cite{ref12} exactly both
the inequality indices $g$ and $k$ here.  We can now
calculate  the Lorenz function
$L(f) = \int_0^m x \rho(x) dx = 1 - (m + 1) exp(-m)$,
where  $f = \int_0^m \rho(x) dx = 1 - exp(-m)$,
giving $m = - ln(1 - f)$, giving  in turn
$L(f) = 1 - (1 - f)[1 - ln(1 - f)]$. One thus gets
(see Fig. 1, noting the area under the equality
line to be 1/2), the Gini index
$g = 1 - 2\int_0^1 L(x)dx = 1/2$  and the Kolkata
index $k$ given by the self-consistent equation
$1 - k = L(k)$ or $1 - 2k = (1 - k)[ln(1 - k)]$,
giving $k \simeq 0.68$.

\textbf{b)} We now proceed to to study numerically the
uniform saving exchange model and the
corresponding Gini and Kolkata  indices.
In this model (called CC model \cite{ref11,II.4,II.5})
we consider again a closed economic system with a
fixed amount of money $M$ and a fixed number of
agents $N= M$, where the agents are interacting
(trading with)  each other  by exchanging their
money. A saving propensity  $\lambda
(0 \le \lambda \le 1)$ of the agents is introduced
in this model, such that during each (two-body)
trade event,  each of the agents saves  a fraction
$\lambda$ of their money in possession  at that
point of time (trade) and the rest of money is
again exchanged randomly between the two trade
partners. The exchange of money $m_i (t)$ between
two traders ($i$ and $j$) at time $t$ can be
expressed as 
\begin{subequations}
\begin{align}
 m_i(t+1)=\lambda m_i(t) + \epsilon_{ij} (1-\lambda)(m_i(t)+m_j(t))\\
 m_j(t+1)=\lambda m_j(t) + \epsilon_{ij} (1-\lambda)(m_i(t)+m_j(t))
\end{align}
\end{subequations}

\begin{figure}[H]
\centering
\includegraphics[width=0.7\textwidth]{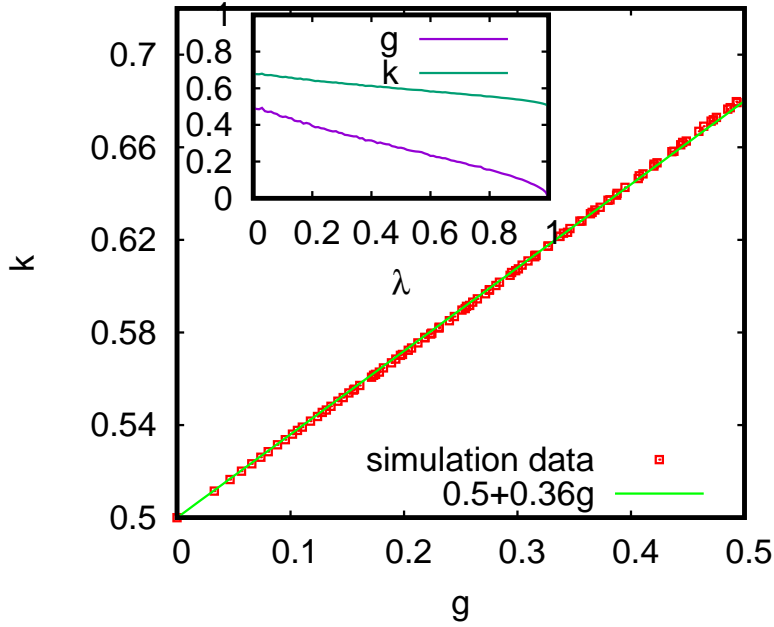}
\caption{Variation of the Kolkata index ($k$) against the Gini
index ($g$ ) for kinetic exchange model with uniform
saving propensity ($\lambda$; CC model). For $\lambda
= 0$ the estimated values of $k \simeq 0.68$ and
$g \simeq 0.5$ conforms the theoretically estimated
exact values of $g = 0.5$ and $k \simeq 0.682$,
discussed above. As expected, with increasing saving
propensity, the inequality decreases and $g$ tends to
vanish and $k$ approaches 0.5 as $\lambda$ tends towards
unity.  The simulation data fit well with the relation
$k = 0.5 + C g$  with $C \simeq 0.36$ (a Landau-like
theory for the Lorenz function, discussed below, gives
$C = 3/8$). Inset shows the variations of $g$ and $k$
with $\lambda$. }
\label{fig0}
\end{figure}


\textbf{c)} We now proceed with an approximate expansion
\cite{ref13} of the Lorenz function $L(p)$,
employing a  Landau-like  argument \cite{ref2} for the
expansion of free energy. A Landau-type minimal
expansion of the Lorenz function $L(f)$ up to
the quadratic term $f$ gives
\begin{equation}
 L(f) = Af + Bf^2, ~ A > 0, B > 0, A + B = 1.
\end{equation}
\noindent As may be noted, the above expansion gives
$L(0) = 0$ and $L(1) = 1$, and with the linear term
alone, the Lorenz function can represent only the
equality line (see Fig. 1). One can now calculate
the Gini index $g = 1 - 2\int^1_0L(f)df$
giving $A = 1- 3g$ and $B = 3g$. The value
of $k$ index can be found from the relation (see Fig.
1) $L(k) = 1 - k$, giving $3gk^2 + (2 - 3g)k -1 = 0$,
or

\begin{equation}
 k = \frac{(3g - 2) \pm \sqrt{(2 - 3g)^2 + 12g}}{6g}. 
\end{equation}

\noindent In the $g \to 0$ limit, the above expression gives \cite{ref13}
$k = 1/2 + (3/8)g$, which suggests $k = g = 0.8$, the
Pareto value under extreme competition \cite{ref8}. Of course,
the full relation (3) gives $g = k \simeq 0. 74$, which is much
smaller than the observed values around 0.86 \cite{ref1,ref10} and
even the Pareto value 0.80 (corresponding to Pareto's 80-20 law \cite{ref8}).

\textbf{d)} We now discuss about the self-organized  appearance
of minimum energy or poverty level \cite{ref12} in the kinetic
exchange model and in its extension with uniform
saving propensity, namely in the CC model. Specifically
we consider here a kinetic exchange model where one of
the agents in the chosen pair is necessarily the poorest
(in money or energy) at that point of time (trade or
scattering) and the other one is randomly chosen from
the rest. Here we vary the saving propensity ($\lambda$)
for values other than 0, and the exchange of money will
follow the same rule as described by equation (3.1). An
important observation is the  spontaneous appearance
of a Self-Organized Poverty (or energy)  Level (SOPL)
in the steady state distribution, below which the
distribution function vanishes ($\rho(m) = 0$). For $\lambda
= 0$ the SOPL  occurs at $m = \theta (\lambda = 0)
\simeq 0.61$ (see Figs. 3 and 4). This SOPL ($\theta
(\lambda)$) also increases  with increasing
values of $\lambda$   (see Fig. 3) and the
$\theta (\lambda)$  approaches unity as  $\lambda$
approaches unity (see Fig. 4).

\begin{figure}[H]
\centering
\includegraphics[width=0.7\textwidth]{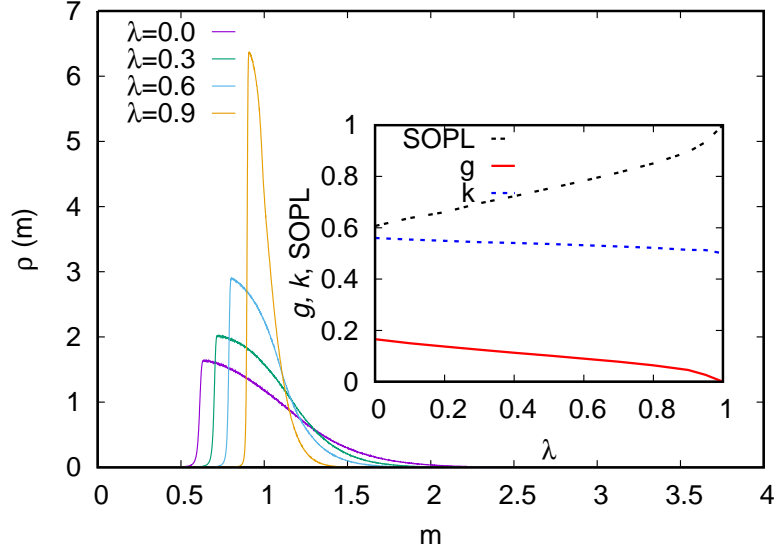}
\caption{Steady state money distribution $\rho(m)$ for fixed saving
propensity $\lambda$  in the kinetic exchange model where in each
two-body scattering (trade) one particle (trader) is  with least energy
(money) the other one is chosen randomly from the rest.
 In the inset the variations of inequality indices
$k$, $g$ and location of the Self-Organized Minimum Energy (Poverty)
Level (SOPL) are shown against fixed saving propensity $\lambda$. Adapted from \cite{ref12}.}
\label{fig-sopl-ba}
\end{figure}

\textbf{e)} We now sketch a mean field like argument to
estimate the value of $\theta$, the Self Organized Poverty (Energy) Level (SOPL). If
we assume, following ref. \cite{ref12}, that generally
the steady state distribution $\rho(m)$ of money or
energy in such kinetic exchange models of SOPL
remains Gibbs-like (exponentially decaying, but
with shifted origin to $m = \theta$:
$\rho(m) = \exp [- (m-\theta)]$ for $m > \theta$
and $\rho(m) = 0$  otherwise), the  average energy
of any of the traders or particles will be
equal to [$\theta + (\theta +1)e^{-1}]/2$, which has
to be greater than $\theta$. This is because, one of the trade partners
must have (with probability 1) $\theta$ amount of money,
while the other can be any one else and can be
assumed to have an average  money ($M/N$ = 1,
shifted by  the minimum $\theta$) and hence
comes with a probability $\exp (- 1)$. Finally,
there will be on average a 50-50 share for any
one and that share value has to be equal to or above
the minimum ($\theta$).  This gives the estimate
$\theta \le [\theta+(1+\theta)e^{-1}]/2$ or $\theta \le e^{-1}/(1 - e^{-1})$, giving $\theta \le
0.58$. This upper bound is somewhat less than the observed value (see Fig. \ref{fig-sopl-ba}; $\theta (0)\simeq0.61$) at $\lambda  = 0$. For
$\lambda$ approaching unity, the distribution
any way approaches equality (at $m = 1$) \cite{ref11,ref12}. Hence the above equation simplifies to
$\theta \le [\theta + 1]/2$, or $\theta \le 1$,
which is clearly observed.

\subsection{Social indices $g$ \& $k$ in  percolation model}
\begin{figure}[H]
\centering
\includegraphics[width=0.7\textwidth]{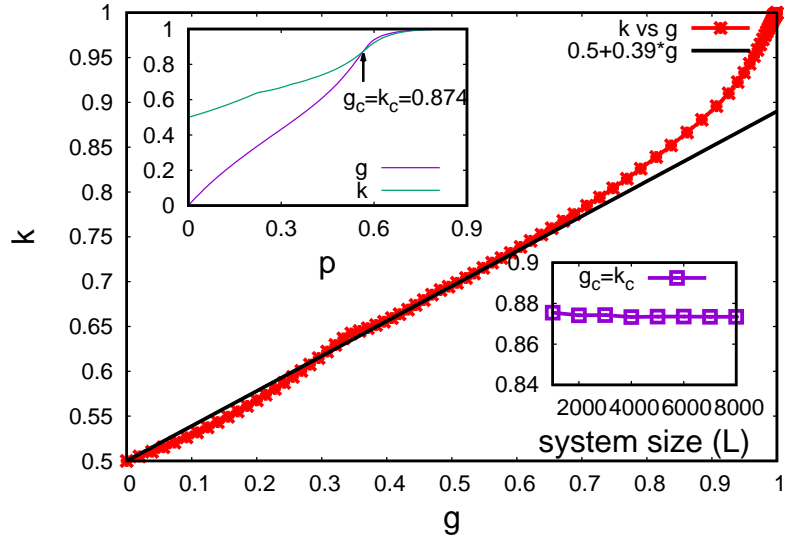}
\caption{Kolkata index ($k$) against Gini index ($g$) for 2-d site  percolation on square lattice is shown here. The initial slope of the simulation data fits well with the relation $k=0.5+0.39g$. The upper inset shows the variations of $g$ and $k$ with occupancy probability $p$ and the crossing value  $k_c$ or
$g_c$  of $k$ and $g$ is about  0.874 and occurs at $p \simeq 0.565$, somewhat
below $p_c \simeq 0.593$.}
\label{fig0-1}
\end{figure}
In percolation models, a regular square lattice ($L \times L$)  with site occupation probability $p$ is considered. The cluster size distribution is measured for different $p$ values. The cluster size $s$ measures the  number of the nearest neighbour occupying sites  and the number $n(s)$ of $s$ size cluster at any particular $p$  will give  the cluster size distribution (at any $p$), which has been utilized to generate the Lorenz function. The inequality indices  $g$ and $k$ are obtained from the Lorenz function (see Fig. 1) for distributed cluster sizes. The Fig. \ref{fig0-1} shows the variation of the Kolkata index ($k$) against the Gini index ($g$) of the cluster sizes for different site occupation probability $p$ (we performed the simulations for system size $4000 \times 4000$). The initial slope of the simulation data fits well with the relation $k = 0.5 + C*g$ ($C \simeq 0.39$) . The upper inset shows the variation of $g$ and $k$ with occupation probability $p$ and the crossing value of the two indices $k_c=g_c \simeq 0.874$ at $p \simeq 0.565$ (while the critical point is $p_c\simeq 0.593$). The lower inset shows the variation $k_c$ or $g_c$ with system size ($L$).

\begin{figure}[tbh]
\centering
\includegraphics[width=0.5\textwidth]{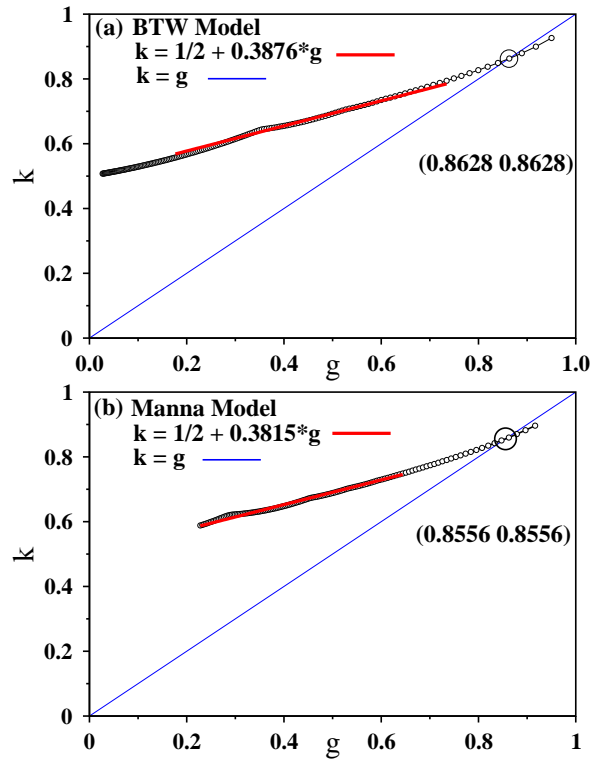}
\caption{The Kolkata index $k$ against Gini index $g$ is plotted for (a)  BTW Model and (b)  Manna Model. The initial slope of the simulation data fits well with the relation $k=0.5+C*g$ ($C=0.385\pm0.005$) and $g_c=k_c=0.860\pm0.005$ is the crossing point with the $g = k$ line. Taken from ref. \cite{ref17}.}
\label{fig4b}
\end{figure}


\subsection{Social indices $g$ \& $k$ in Bak-Tang-Wiesenfeld (BTW), Manna sandpile and centrally pulled Fiber Bundle models}
{\bf a)} Inequality in the Bak-Tang-Wiesenfeld (BTW) and Manna sandpile models: In BTW model on a square lattice, the sand grains are added one by one at randomly selected sites.  The heights of the sand columns at different sizes will increase by addition of these sand grains. When the sand column height ($h$) at any site reaches a threshold value (4, in the BTW model), the column becomes unstable and  the sand grains  from the unstable sites are equally shared among the  neighboring (4) sites uniformly. This may cause the neighboring columns to become unstable and the avalanche continues. In the Manna sandpile model (on square lattice again) when the sand column height reaches a threshold value 2, the column becomes unstable and the sand grains from the unstable column will be shared randomly by two of the neighbouring columns which may become unstable again and the avalanche may continue. Therefore an avalanche of topplings will occur in both the models  until height $h$ at all the lattice sites become less than the respective threshold values. The random addition of sand grains to the sand pile then induces further dynamics in the models. The avalanche size $s$ measures the total number of toppings in one go, without any further addition of sand grain to the system and the number of $s$ size avalanches $n(s)$ in  the steady state of the models will give respective avalanche size distributions, which have been utilized to generate the respective Lorenz functions.

The inequality indices ($g$ and $k$) are obtained from the above mentioned Lorenz functions for the respective models. The Figs. \ref{fig4b}(a) and  \ref{fig4b}(b) show the variations of Kolkata index against Gini index for different average heights of the sand columns in the BTW and Manna models respectively. The initial variations of the simulation data in both the models seem to fit well with the relation $k=0.5+C*g$ ($C=0.385\pm0.005$) and $g_c=k_c=0.860\pm0.005$ for the crossing point. It may be mentioned that this crossing of $g$ and $k$ occurs at the values of average height $h\simeq 2.087$ (slightly below the actual critical height $h_c=17/8$ for the BTW model) and at $h=0.6859$ (compared to the actual critical height $h_c\simeq 0.7172$ for the Manna model) \cite{ref15}.

{\bf b)} Inequality in centrally pulled fiber bundle model: In this version of the fiber bundle model \cite{biswas_13}, initially a load is applied only at a centrally located fiber in a two dimensional arrangement of fibers. The applied load (pull) is slowly increased at a constant rate. When that fiber breaks, the load is shared equally between its nearest neighbors. Should some of those neighbors fail, their load would be equally redistributed among all fibers that have at least one broken neighbor. The redistribution process occurs at a much faster rate than the pulling. Hence, the load per fiber value along the centrally located damage boundary fluctuates around a steady state. The numbers of fiber broken in an avalanche show power law size distribution. Unlike the usual version of the fiber bundle, where the dynamics eventually stops due to a catastrophic failure of the whole system, in this case it continues, until the damage boundary reaches the system boundary, i.e. in the thermodynamic limit of infinite system size, the dynamics keeps on going. 

The inequality of the avalanches could be measured in the same way as in the case of the SOC models mentioned above. The plots of $g$ vs $k$ are shown in Fig. \ref{fbm_soc_gk_compare}. 

For a theoretical understanding of this behavior, if the Lorenz function is written as $L(f)=f^{\theta}$, then it is known that
$g=(\theta-1)/(\theta+1)$. In other words, then $\theta=(1+g)/(1-g).$
Now, one gets $k$ from solving  $1-k=k^{\theta}$
Then clearly,
\begin{equation}
g=\frac{ln(1-k)-ln (k)}{ln (1-k)+ ln (k)}.
\label{gk_relation}
\end{equation}
This relation does not restrict the values of $g$ and $k$ and should be valid as long as the form of the Lorenz function is a power-law. It can be compared
with the numerical observation of the relation between $g$ and $k$ in the SOC models (discussed later). If, however, one concentrates in the limit of small values of $g$ and $k$, i.e., $k=1/2+\epsilon$, where $\epsilon \to 0$, then 
the above relation reduces to
\begin{equation}
 k=\frac{1}{2}+\frac{ln (2)}{2}g.
\end{equation}
The relation Eq. (\ref{gk_relation}) is compared with the simulation of centrally loaded FBM in Fig. \ref{fbm_soc_gk_compare}.

\begin{figure}[H]
\begin {center}
\includegraphics[width=0.95\textwidth]{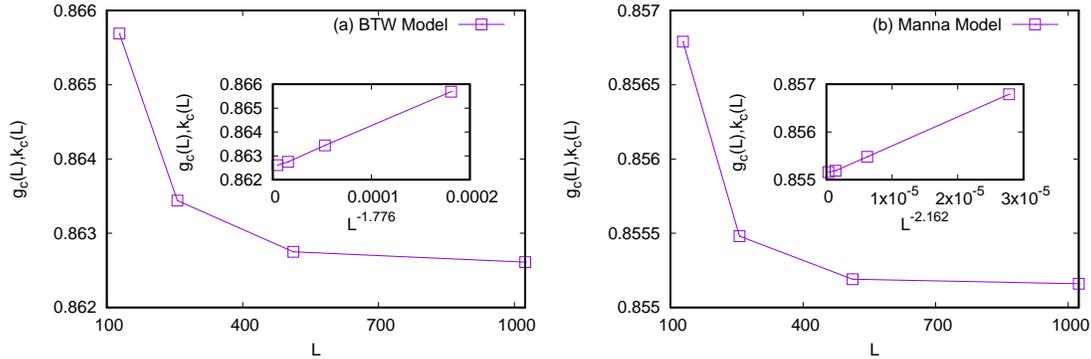}
\end {center}
\caption{(a)  BTW sandpile model:
 The values of the Kolkata  $k_c(L)$ and the Gini  $g_c(L)$ indices (at the point when they cross) have been plotted
    as a function of the system sizes $L$ = 128, 256, 512 and 1024. The crossing point ($g=k=g_c=k_c$) values of the indices
     decrease with the system size.
(Inset) The values of $k_c=g_c$ as function of $L$ have been extrapolated with respect to $L^{-1/\nu}$ where the value of $\nu$ for BTW model has
    been tuned for the best possible linear fit of the data and it gives $1/\nu=1.776$ which in turn  $k_c = g_c = 0.863$ in the limit of $L\rightarrow \infty$. (b) Manna sandpile model: The values of the Kolkata index $k_c(L)$ and the Gini index $g_c(L)$ have been plotted
    against the system sizes $L$ = 128, 256, 512 and 1024. The values of the indices
     decrease with the system size.
(Inset) The values of $k_c=g_c$ as function of $L$ have been extrapolated with respect to $L^{-1/\nu}$ and value of the exponent $\nu$ has
    been tuned for the best possible fit of the data. The plot shows that the best fit corresponds
    to $1/\nu=2.162$ for the Manna model, giving in turn $k_c = g_c = 0.8554$ in the limit of $L\rightarrow \infty$. Adapted from ref. \cite{ref17}.
}
\label {FIG03}
\end{figure}


\begin{figure}[H]
\begin {center}
\includegraphics[width=0.5\textwidth]{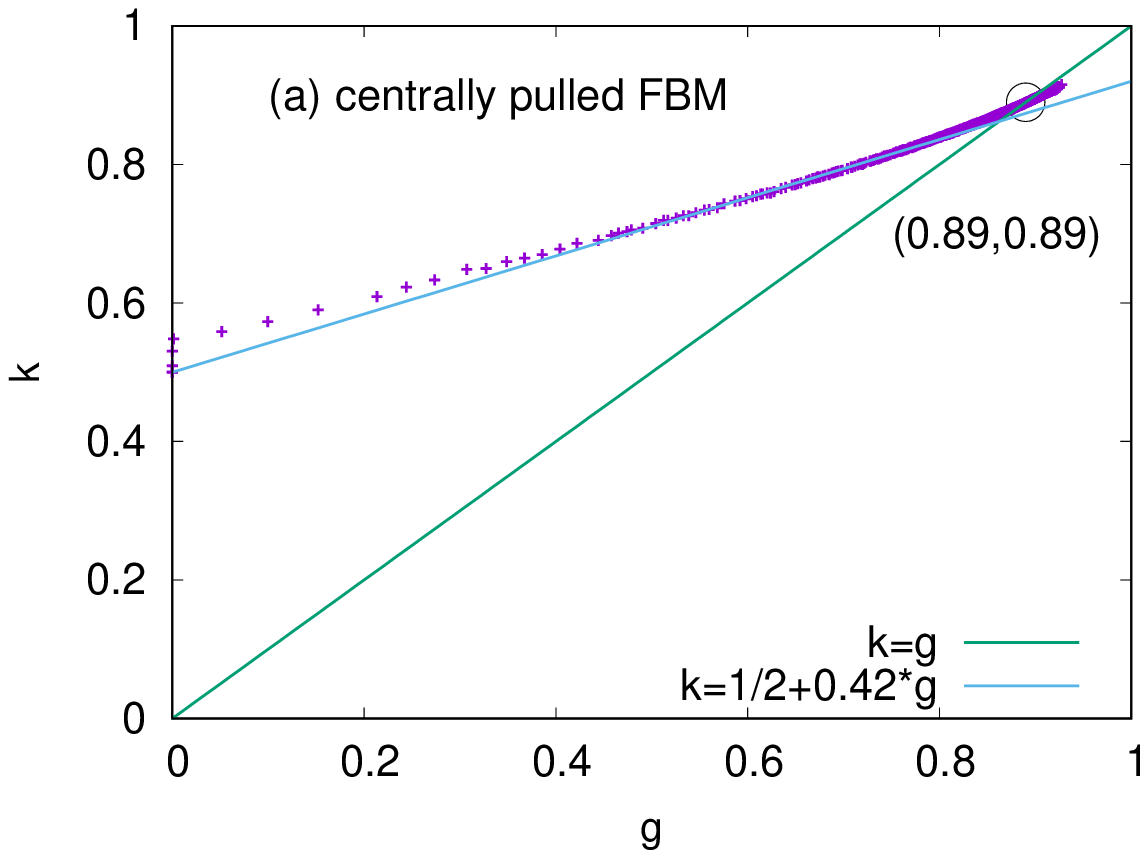}
\includegraphics[width=0.5\textwidth]{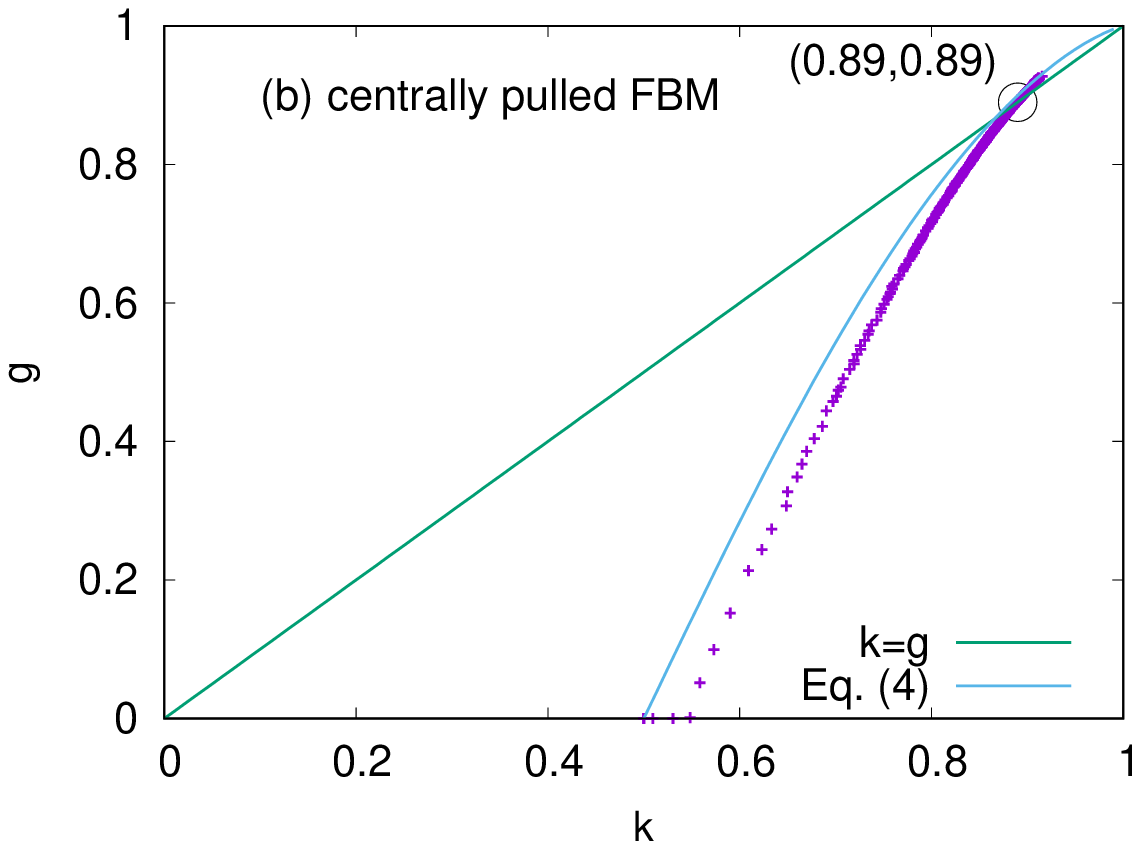}
\end {center}
\caption{The variations of $k$ against $g$ indices are shown for  the centrally pulled fiber bundle model. (a) The line is a fit for the initial growth of $g$ and $k$, while the $g=k$ line is also shown for a guide to the eye.  Adapted from ref. \cite{ref17}. (b) The same variation is plotted with $g$ against $k$ and compared with Eq. (\ref{gk_relation}).}
\label{fbm_soc_gk_compare}
\end{figure}

\subsection{Social indices $g$ \& $k$ in the Fiber Bundle Model}
In the previous subsection IIC (b), we considered a self-organized
version of the
fiber bundle model, where the breaking  dynamics continues indefinitely  as the
active fiber bundle system (on the periphery of the central broken patch)  grows
continuously in size as the central pull or load  is increasing with time.

In the standard version of the fiber bundle model (see e.g.,  \cite{ref16,ref17,ref18})
of course, the
breaking dynamics stops as the entire stem (fixed in size) fails. Here
the collective
dynamics  of failure or breaking in any non-brittle material sample proceeds
through the failures of individual elements of the material,  as
the external load or stress on the sample increases.
The subsequent redistribution of the load shares among
the surviving elements and consequent further failures
and avalanches (even when the external load does not
increase any further).  These bursts of elastic
energy released (experimentally detected as acoustic
emissions) until the complete breakdown of the
material, are widely studied  \cite{ref11}. These bursts or
avalanches are also studied often in models, like
the Fiber Bundle Model (FBM) \cite{ref12,ref13},  both
analytically and numerically. An avalanche is
defined as the size or mass of the failure  events
taking place in the system in going from one
stable state to the next, as the external load on
the system is increased further to trigger a failure
activity (load gradually increased) while the
(relatively faster) internal dynamics continues due
to load redistribution among the surviving fibers
and consequent failures due to such increased load
on them.  The avalanche size could also be
measured by the amount of elastic energy released
from these failed elements. Its distribution would
then correspond more naturally to the elastic
energy emissions. In the mean-field version of the model considered here, these two quantities (avalanche and energy) have the same size distribution function. For simplicity,  we will
consider here the avalanche size to be given
only by the number of failed elements.  For
successive increases in the external load, further
avalanches of different sizes occur with various
frequencies. As in the wealth distributions, the
distributions of the avalanche sizes, across a
broad class of systems, show the common feature
of having relatively larger number of smaller
events (poorer people) and much fewer number of
large ones (richer people), indicating similar
nonlinear nature of the corresponding Lorenz
function $L(f)$ (see Fig. 1). In statistical
physics, however, we usually concentrate on the
(fractal) structure the biggest avalanche size,
which becomes of the order of the system size and
causes the eventual macroscopic failure of the
sample (see e.g., \cite{ref3,ref4}). Some recent studies
\cite{ref15,ref19} on the social inequality indices in
equal-load-sharing FBMs \cite{ref16,ref17,ref18} having widely
different fiber breaking threshold distributions,
suggest gradual increase of the Gini and Kolkata
indices with increasing load on the bundles,
towards some universal terminal values, namely
$g \simeq 0.45$ and $k \simeq  0.65$ respectively,
at the breaking point $\sigma_c$ (breaking load
per fiber) of the respective bundles. Needless
to mention, monitoring the values of such (social)
indices $g$ and $k$ for failure avalanches (usually measured as acoustic emissions) can
therefore  provide an easy and unique precursor
signal \cite{ref15,ref19} to the imminent disasters.

Indeed, our recent analysis \cite{in_prep} of USGS earthquake magnitude data for 22 years (2000-21) show universal social inequality indices terminal values.
For fiber bundle model, an analytical estimate of the failure point values of $g$ and $k$ for particular limits (equally spaced failure thresholds and
equal load increase) can be attempted \cite{in_prep}. It can give an idea of why the limiting values are independent of the different threshold distributions.

\begin{figure}[H]
\centering
\includegraphics[width=7cm]{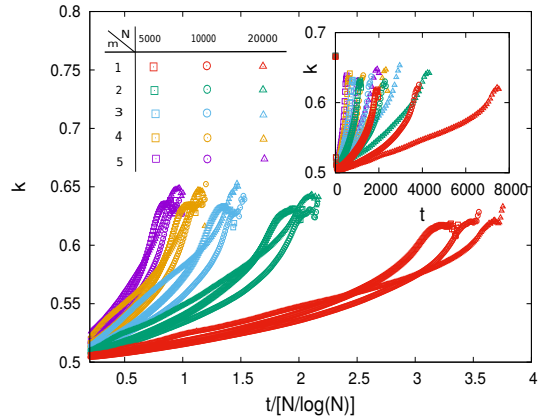}
\caption{The Kolkata index $k$ for the avalanche
distribution $D(\Delta)$ as the dynamics of failure continues in the FBM (in
the ELS scheme), where the individual fiber thresholds drawn from a
Weibull distribution. The estimated values of the
index $k$  at different times $t$ (scaled by $N/log N$) are plotted until
complete failure of the bundle (with  disorder characterized  by different
 Weibull  moduli ($m$) indicated using different colors). The terminal
value of the
$k$-index, prior to complete failure bundle, seems to reach a threshold ($0.62
\pm 0.03$) and this terminal value is weakly dependent on $m$.  The inset
shows the
variations of $k$ index with unscaled time. Taken from \cite{ref19}.}
\label{kindex_weibull}
\end{figure}

\begin{figure}[H]
\centering
\includegraphics[width=14cm]{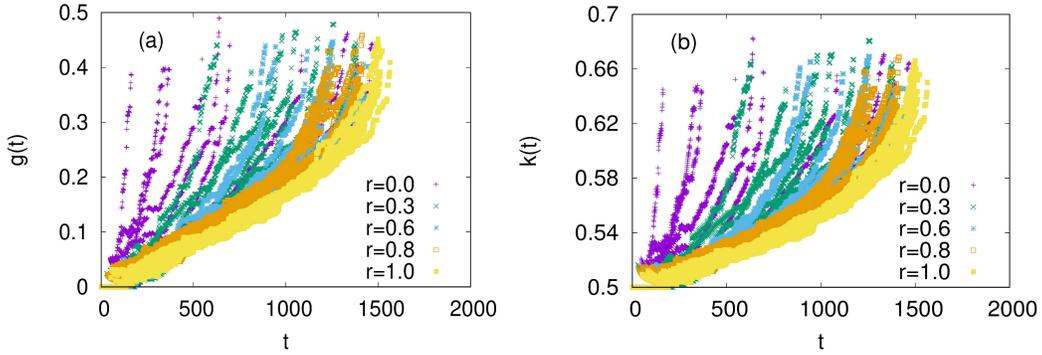}
\caption{The time variations of (a) $g(t)$ and (b) $k(t)$ are shown when individual
samples are of sizes between $L_{min}$ and $L_{max}$ with uniform probability, where $r=L_{min}/L_{max}$ and 
10 time series are shown for each value of $r$. While the failure times for the individual samples are vastly 
different, the terminal values of $g=g_f$ and $k=k_f$ are narrowly distributed. The
failure thresholds are taken from a Weibull distribution with a shape parameter value $\kappa=3$ and $L_{max}=10000$. Adapted from \cite{ref20}. }
\label{fig_s2}
\end{figure}


\section{Summary and discussion}
\label{sec3}
Many physical systems close to their critical points exhibit large fluctuations. In spite of many differences, large groups of systems show 
universal nature in the statistical features of such fluctuations. In other words, such differences  are
irrelevant in the renormalization group sense, and the critical points are characterized by sets of critical exponents that only depend on a few subtle
parameters (space dimension, order-parameter dimension, etc.). Nevertheless, in measuring the critical exponents, the critical points need to be known, which
can depend on many details of the particular system under investigation. There can also be some situations where the system can only be probed from one side
of the critical point (e.g., breakdown of driven disordered solids). In such cases, knowledge of the proximity to the critical point (imminent breakdown) is 
often crucial. Knowing some typical universal values of the critical exponents alone does not help in determining the proximity to the critical point. 

Here we have outlined, in a variety of physical systems, how the characterization of the (social) inequality in the response statistics of systems close to
the critical points can help in determining the proximity to such a point. It is remarkable that the signals that the inequality measures (Gini and Kolkata indices) give,
are quite universal and independent of the value of the critical point. Therefore, it can serve as a useful indicator to an imminent critical point, just from the fluctuations of the order parameters and without requiring the knowledge of the specific value of a critical point.

\begin{figure}[tbh]
\centering
\includegraphics[width=7cm]{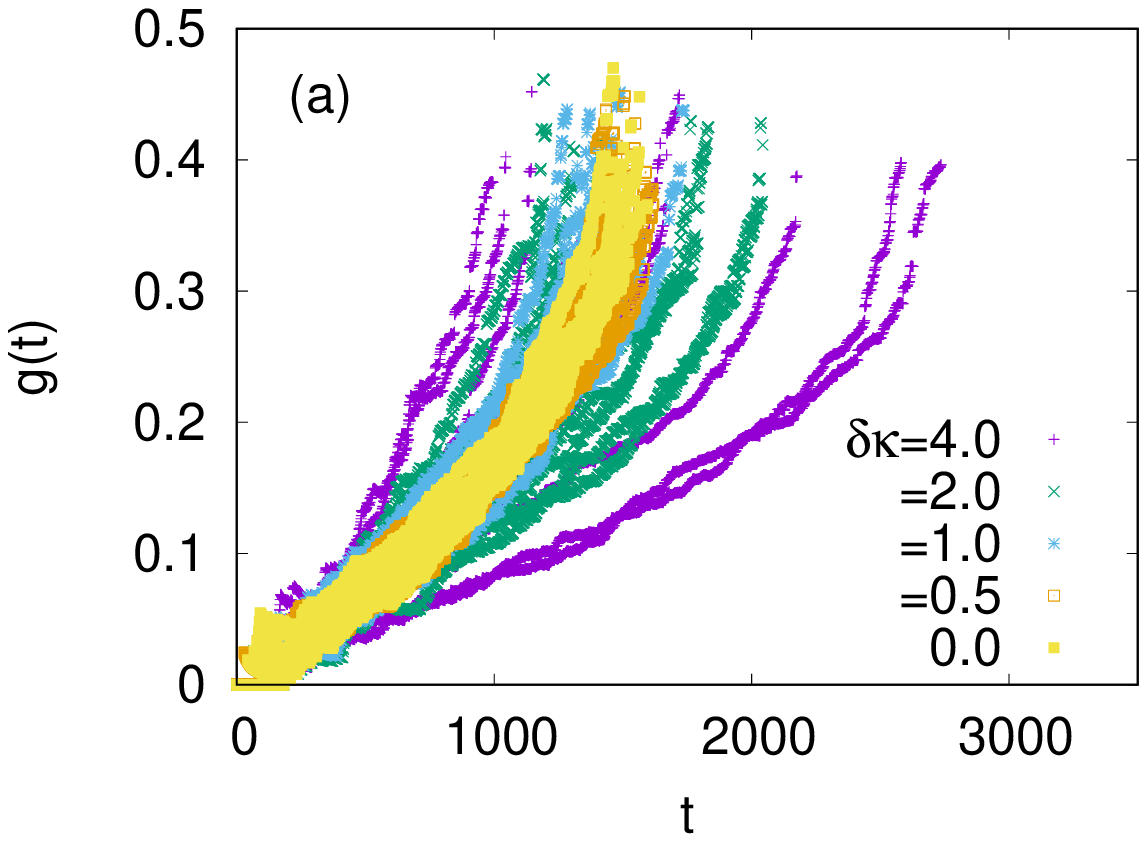}
\includegraphics[width=7cm]{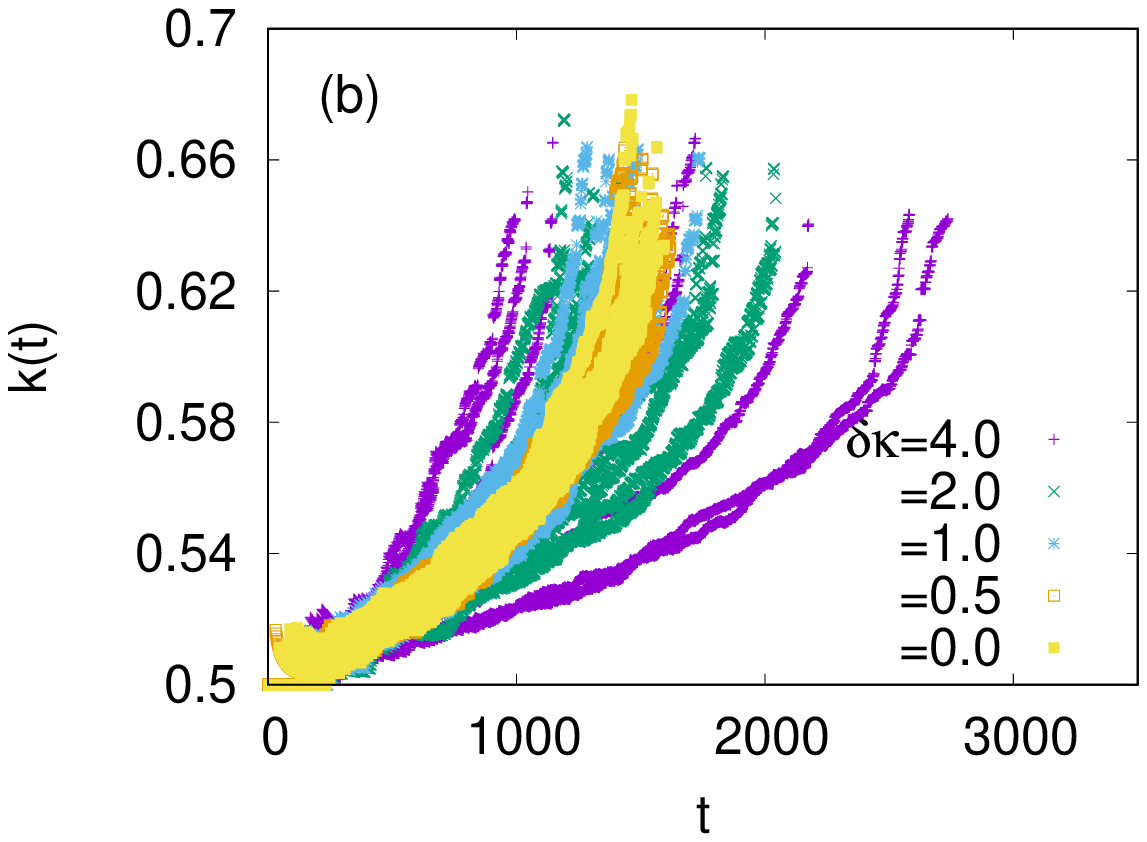}
\caption{The time variations of (a) $g(t)$ and (b) $k(t)$ are shown when individual
samples are of different disorder strengths -- Weibull threshold distributions with 
shape parameters distributed uniformly between $\kappa_{min}$ and $\kappa_{max}$, with 10
samples for each $\kappa_{min},\kappa_{max}$ pair. In the labels, $\delta \kappa=\kappa_{max}-\kappa_{min}$, with $(\kappa_{max}+\kappa_{min})/2=3.0$, in each case.  The system size is 1000 always. Again, the 
failure times are vastly different, but the terminal values of $g=g_f$ and $k=k_f$ are narrowly 
distributed. Adapted from \cite{ref20}. }
\label{fig_s3}
\end{figure}

\begin{figure}[tbh]
\centering
\includegraphics[width=9cm]{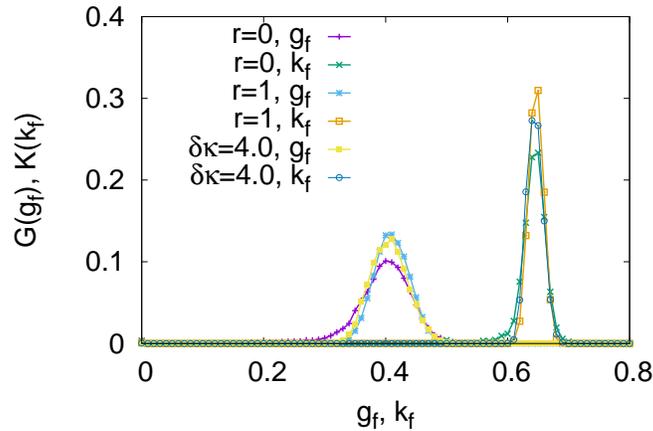}
\caption{The probability distributions of the terminal values of $g=g_f$ (denoted by $G(g_f)$) and $k=k_f$ (denoted by $K(k_f)$) are shown for extreme values of $r=L_{min}/L_{max}=0,1$ and 
$\delta\kappa=4.0$. The distributions show peak near $\langle g_f\rangle=0.41\pm 0.04$ and $\langle k_f\rangle=0.64\pm 0.02$. The averages are done over 10000 ensembles. Adapted from \cite{ref20}.}
\label{fig_s3.5}
\end{figure}

We have analyzed here the kinetic wealth exchange model, geometrical percolation on two dimensional lattice, self-organized critical models and the fiber bundle model of failure in disordered solids.
Specifically, in sec. IIA we have discussed the kinetic wealth exchange model and the appearance of the self-organized poverty line and the variations of the inequality indices with an analytical estimate using a Landau-like expansion of the Lorenz function. In sec. IIB, inequality indices are computed from the unequal distributions of clusters (occupied nearest neighbor sites) on a square lattice for different values of occupation probability. The crossing of $g$ and $k$ occurs ($k_c=g_c\simeq 0.86$) at an occupation probability just below the critical (percolation) probability.  In sec. IIC, self-organized critical dynamics in sandpile (BTW and Manna) and centrally pulled fiber bundle models are studied in terms of the inequalities in their avalanche statistics. As before, the crossing point of the inequality indices $g$ and $k$ ($k_c=g_c\simeq 0.86$) again  indicates proximity to the onset of self-organized critical state. Finally, in sec. IID, the inequalities in avalanches are discussed for the fiber bundle model where the dynamics terminates at a catastrophic failure point, unlike the self-organized dynamical state discussed in sec. IIC. In this case, the inequality indices do not cross, but the terminal values are broadly universal ($g_f\simeq 0.45$ and $k_f\simeq 0.65$) and therefore could be useful in predicting the imminent failure point.

Except in the last case, the fluctuations in the other models (that of wealth of an individual, sites in the largest connected cluster or grains in an avalanche event) are only limited by the system size (or at least an increasing function of the same). This so called `unrestricted competition' leads to remarkably robust characterizations of the inequality measures. Particularly, in spite of the wide variety of the physical systems considered here, in their dynamics, dimensionality and consequently the universality classes, the inequality indices Gini ($g$) and Kolkata ($k$) cross at $g_c=k_c\simeq0.86$, in almost all cases (within a limited range of deviation) just preceding the critical point. In the case of the fiber bundle model of catastrophic failure the dynamics stops. In such cases, therefore,  $g$ and $k$ do not cross, but nevertheless show robust feature (with respect to disorder distribution, system sizes) in terms of their values ($g_f\simeq 0.45$ and $k_f\simeq 0.65$) at the failure point \cite{ref19,ref20} and seems to support also the observations from analysis of earthquake data \cite{in_prep}.

Analytical understanding of these features are still lacking. However, we have discussed here (in Sec. IIA (c)) the minimal Landau-like expansion of a Lorenz function that can 
correctly predict the precise relationship between $g$ and $k$ in the small-value limit of $g$ (giving  $k_c=g_c=0.80$, the Pareto value; somewhat less than the observed value $k_c=g_c\simeq 0.86$). 

As can be guessed, a robust measure indicating an imminent critical point in a system can be of vital use. We would like to highlight that the social inequality indices are 
extremely promising candidates for such an early signal and indicator for approaching the critical point or the imminent failure in a wide range of physical systems.   

Acknowledgement: We are grateful to our
colleagues Suchismita Banerjee, Nachiketa
Chattophadhyay, Diksha, Jun-ichi Inoue, Bijin
Joseph, Bikash Mandal, Subhrangshu Sekhar Manna,
Manipushpak Mitra, Suresh Mutuswami and
Sanjukta Paul for their collaborations at
various stages of the development of this
study. BKC is grateful to the Indian National
Science Academy for their Senior Scientist
Research Grant.

\end{document}